\documentclass[aps,prl,twocolumn,amsmath,amssymb,showpacs,superscriptaddress,notitlepage,longbibliography]{revtex4-1}

\usepackage[colorlinks=true,linkcolor=blue,anchorcolor=red,citecolor=blue, urlcolor=blue]{hyperref}
\usepackage{bm}
\usepackage{graphicx}
\usepackage{color}
\usepackage{float}
\usepackage{cases}

\begin{document}

\title{Intrinsic spin-orbit torque mechanism for deterministic all-electric switching of noncollinear antiferromagnets}

\author{Yiyuan Chen$^\dag$}
\affiliation{Shenzhen Institute for Quantum Science and Engineering and Department of Physics, Southern University of Science and Technology (SUSTech), Shenzhen 518055, China}
\affiliation{Quantum Science Center of Guangdong-Hong Kong-Macao Greater Bay Area (Guangdong), Shenzhen 518045, China}
\affiliation{Shenzhen Key Laboratory of Quantum Science and Engineering, Shenzhen 518055, China}
\affiliation{International Quantum Academy, Shenzhen 518048, China}

\author{Z. Z. Du$^\dag$}
\affiliation{Shenzhen Institute for Quantum Science and Engineering and Department of Physics, Southern University of Science and Technology (SUSTech), Shenzhen 518055, China}
\affiliation{Shenzhen Key Laboratory of Quantum Science and Engineering, Shenzhen 518055, China}
\affiliation{International Quantum Academy, Shenzhen 518048, China}

\author{Hai-Zhou Lu}
\email{Corresponding author: luhz@sustech.edu.cn}
\affiliation{Shenzhen Institute for Quantum Science and Engineering and Department of Physics, Southern University of Science and Technology (SUSTech), Shenzhen 518055, China}
\affiliation{Quantum Science Center of Guangdong-Hong Kong-Macao Greater Bay Area (Guangdong), Shenzhen 518045, China}
\affiliation{Shenzhen Key Laboratory of Quantum Science and Engineering, Shenzhen 518055, China}
\affiliation{International Quantum Academy, Shenzhen 518048, China}

\author{X. C. Xie}
\affiliation{International Center for Quantum Materials, School of Physics, Peking University, Beijing100871, China}
\affiliation{Institute for Nanoelectronic Devices and Quantum Computing, Fudan University, Shanghai 200433, China}
\affiliation{Hefei National Laboratory, Hefei 230088, China}

\begin{abstract}
Using a pure electric current to control kagome noncollinear antiferromagnets is promising in information storage and processing, but a full description
is still lacking, in particular, on intrinsic (i.e., no external magnetic fields or external spin currents) spin-orbit torques.
In this work, we self-consistently describe the relations among the electronic structure, magnetic structure, spin accumulations, and intrinsic spin-orbit torques, in the magnetic dynamics
of a noncollinear antiferromagnet driven by a pure electric current. Our calculation can yield a critical current density comparable with those in the experiments, when considering the boost from the out-of-plane magnetic dynamics induced by the current-driven spin accumulation on individual magnetic moments. We stress the parity symmetry breaking in deterministic switching among magnetic structures.
This work will be helpful for future applications of noncollinear antiferromagnets.
\end{abstract}

\maketitle

\begin{figure}[h]
\includegraphics[width=0.48\textwidth]{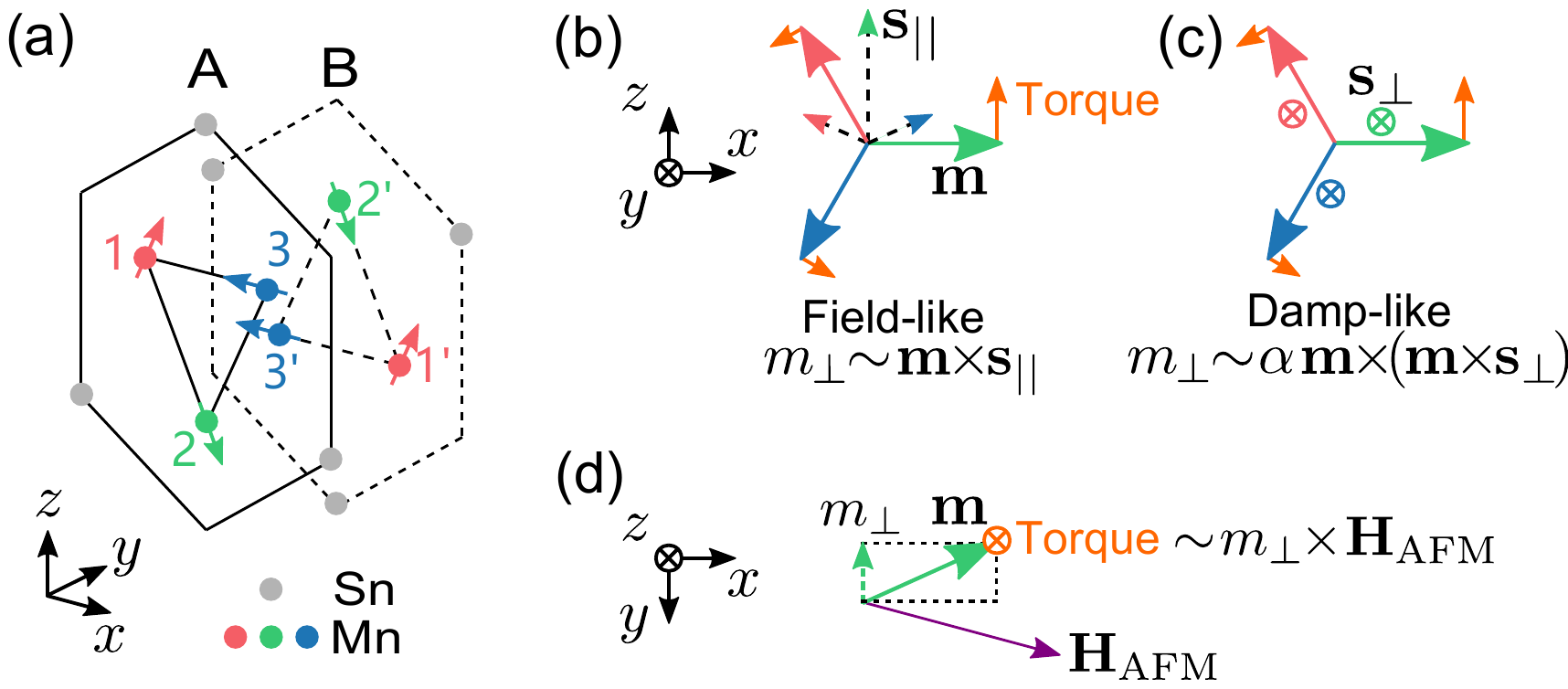}
\caption{(a) A kagome noncollinear antiferromagnet, e.g., Mn$_3$Sn. The red, green, and blue arrows stand for the three Mn moments $\mathbf{m}$ { (1-3 on layer A, 1$'$-3$'$ on layer B)} in the unit cell of a single kagome layer. [(b) and (c)] 
The spin-orbit coupling converts a pure electric current into spin accumulations (dashed arrows for in-plane $\mathbf{s}_{||}$ and $\otimes$ for out-of-plane $\mathbf{s}_{\perp}$). The spin accumulations induce an out-of-plane component $m_\perp$ of $\mathbf{m}$, via the field-like torque $\mathbf{m}\times \mathbf{s}_{||}$ for $\mathbf{s}_{||}$ in (b) or via the damp-like torque $\alpha \mathbf{m}\times (\mathbf{m}\times \mathbf{s}_{\perp})$ for $\mathbf{s}_{\perp}$ in (c). (d) For a moment $\mathbf{m}$, its antiferromagnetic interaction with other two moments acts like an effective field $\mathbf{H}_\mathrm{AFM}$ and $m_\perp \times \mathbf{H}_\mathrm{AFM}$ gives the torques (orange arrows and $\otimes$) that rotate the magnetic structure in the kagome ($x$-$z$) plane. The field-like torque in (b) is dominant, because of the small damping constant $\alpha$ in front of the damp-like torque in (c).}
\label{Fig:Mn3Sn}
\end{figure}

\textcolor{blue}{ \emph{Introduction.}} -
antiferromagnets have been expected to replace the ferromagnets in information storage and processing, because they are robust against perturbations guaranteed by their zero stray field and have higher energy scale for ultrafast performance. However, it is hard to read out ordinary $PT$-symmetric antiferromagnets (e.g., $\uparrow \downarrow \uparrow\downarrow...$) because of their weak magnetization. Recently, kagome noncollinear antiferromagnets [Fig. \ref{Fig:Mn3Sn}(a)] have attracted increasing attention, because they have strong anomalous Hall, Nernst, and magneto-optical Kerr effects \cite{Nakatsuji2015nature,Ikhlas2017naturephysics,li2017prl,Chen2014prl,Kiyohara2016pra,Nayak2016SciAdv,Liu2017SciRep,Liu2018NatureElectron,Ikeda2018apl,Higo2018apl,Higo2018naturephotonics} to serve as readout signals.

Using only pure electric current to manipulate the kagome antiferromagnets can greatly facilitate device applications \cite{Deng2022nsr}. Nevertheless, so far the theories mainly focus on their anomalous \cite{Nagaosa2001prl,Chen2014prl,ZhangY2017prb,Liu2017prl,Liu2018NatureElectron,Batista2020prb} and spin \cite{BHYan2017prl,Kimata2019nature} Hall effects, and their magnetic dynamics driven by the extrinsic mechanisms, i.e., through an extrinsic spin current assisted by a magnetic field \cite{Tsai2020nature,Pal22sa, Xie2022nc}. A theory that fully describes the magnetic dynamics driven by a pure electric current is still lacking, in particular, on their intrinsic spin-orbit torques.

\begin{figure}[htbp]
\includegraphics[width=0.4\textwidth]{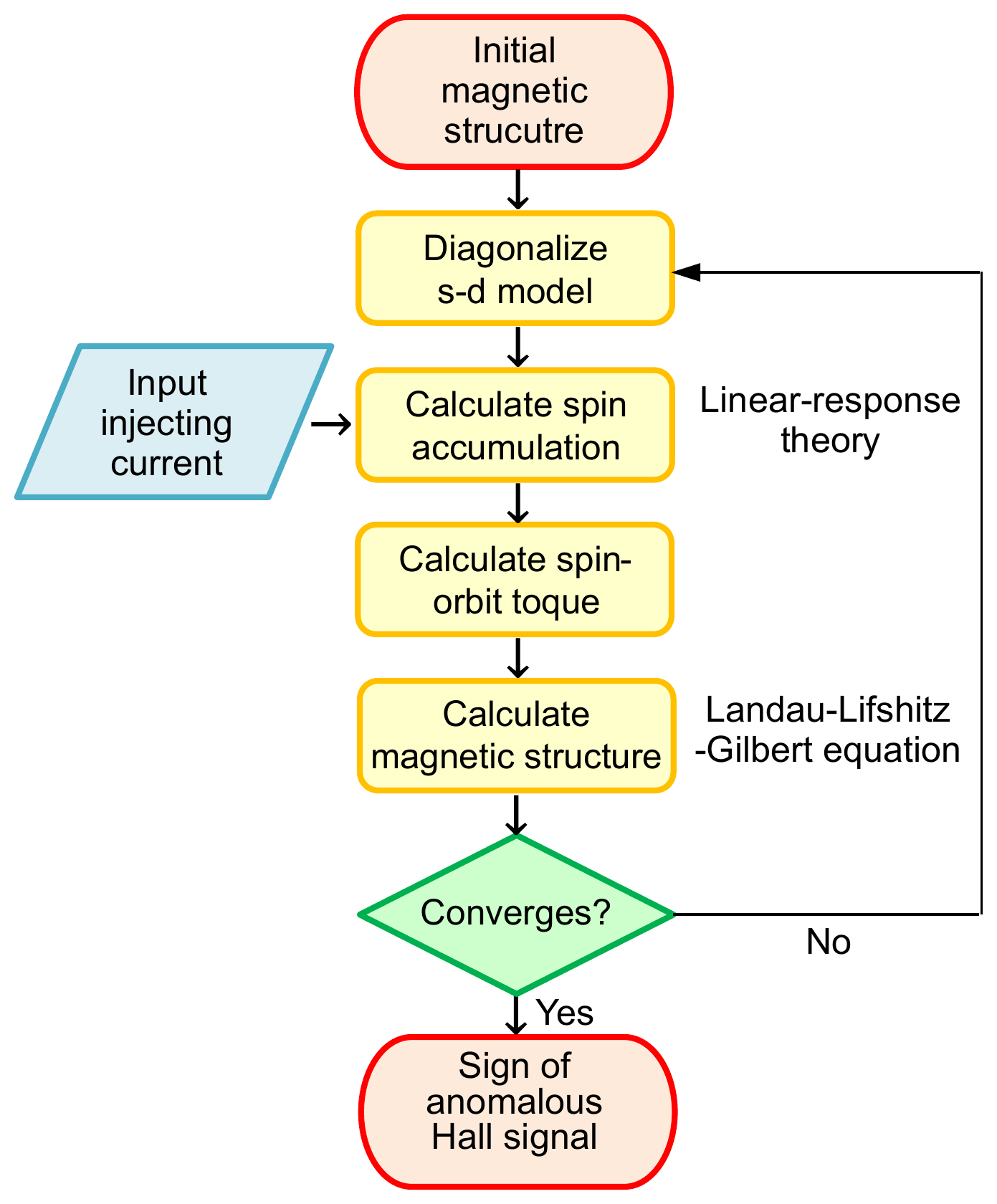}
\caption{The flowchart of the numerical simulation. We start by assuming an initial magnetic structure described by Eq. (\ref{MH}), which serves as local magnetic fields in the $s$-$d$ model Eq. (\ref{Eq:sd}). Then, we use Eqs. (\ref{Eq:Kubo-d}) and (\ref{Eq:Kubo-od}) to calculate the spin accumulations on the Mn atoms at a given injecting current Eq. (\ref{Eq:current}) and use Eq. (\ref{SOT-H}) to calculate the magnetic torques of the spin accumulations in the Landau-Lifshitz-Gilbert equation Eq. (\ref{Eq:LLG}), which generates a new magnetic structure. The steps are iterated until the magnetic structure converges, then the sign of the Hall signal can be determined by $\varphi_2$.
}
\label{Fig:flowchart}
\end{figure}

\begin{figure}[tbp]
\includegraphics[width=0.4\textwidth]{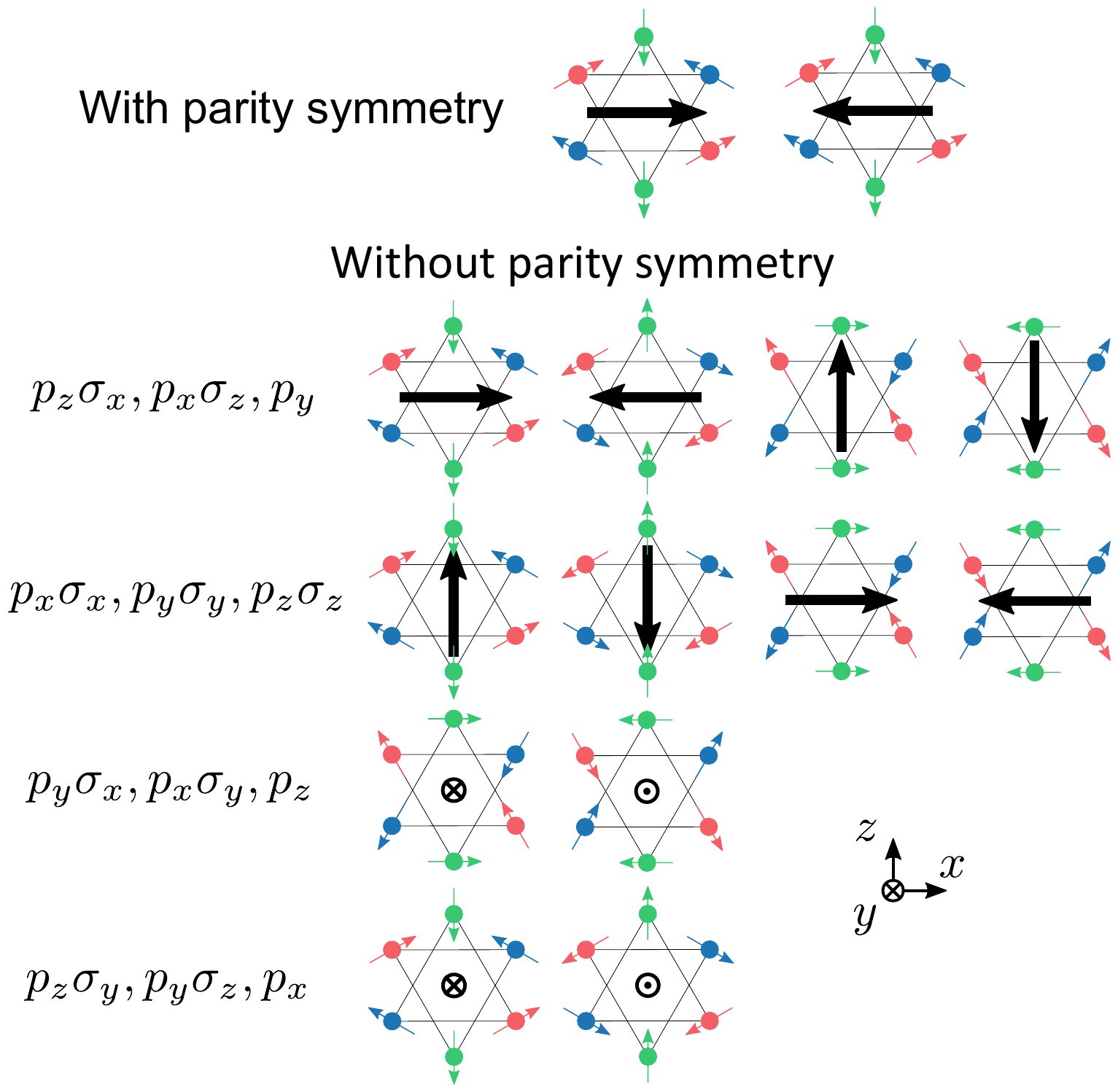}
\caption{Parity symmetry has to be broken for an electric current to deterministically manipulate the Mn moments, because the same magnetic structure is allowed for opposite electric currents (black arrow) under parity symmetry. We list all possible linear-momentum Hamiltonian terms that break parity symmetry and their resulting magnetic structures under the switching current.}
\label{Fig:parity}
\end{figure}

\begin{figure}[htbp]
\includegraphics[width=0.5\textwidth]{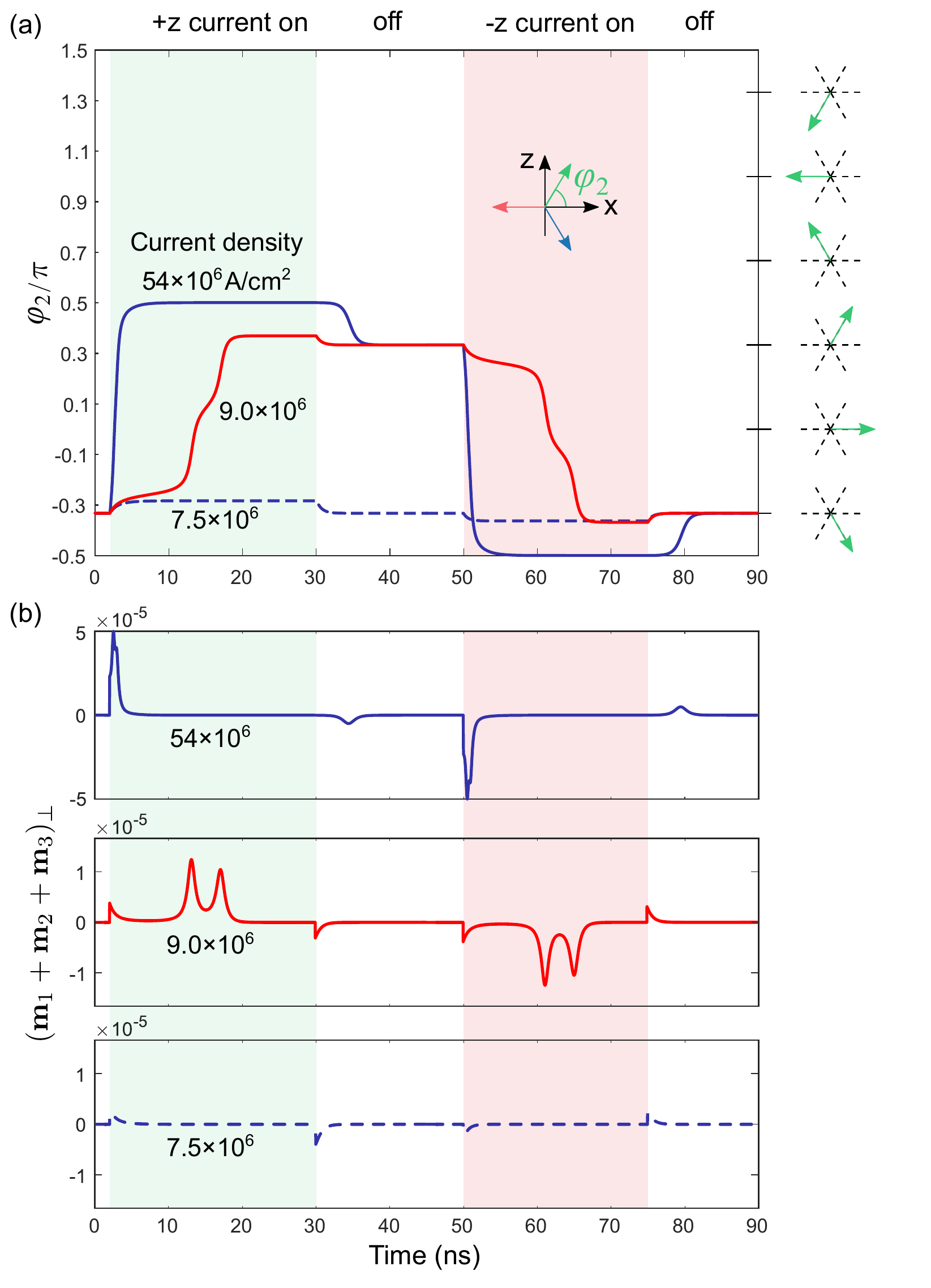}
\caption{(a) Simulated magnetic dynamics of Mn$_3$Sn in terms of the moment angle $\varphi_2$ (inset) of the Mn atom 2,
driven by a pure electric current below ($7.5\times 10^6$), well above ($54\times 10^6$), and at the critical current density ($j\sim 9.0\times 10^6$ A/cm$^2$, comparable with those in the experiment \cite{Deng2022nsr}), for the Weyl model ($\Delta_p^{\mathrm{I}}=0.075$ eV, $\Delta_p^{\mathrm{II}}=0$) in Eq. (\ref{Eq:Weyl}); 
From 2 through 30 ns (from 50 through 75 ns), $j$ is along the $+z$ ($-z$) direction. In absence of $j$, $\varphi_2$ tends to relax at one of the six stable positions marked on the right axis. (b) The total out-of-plane magnetic moment of all three Mn atoms $(\mathbf{m}_1+\mathbf{m}_2+\mathbf{m}_3)_\perp$ for the three current densities in (a). The parameters are $t$=0.25 eV, $J_{sd}$=0.125 eV, $\alpha_1=\alpha_2=\pi/5$, $\theta=\pi/4$ \cite{Batista2020prb}, $\tau=\hbar/2\Gamma$ with $\Gamma$=1.25 meV, $m^*=9.1\times10^{-31}$ kg, $n_{3D}=6\times10^{23}/\mathrm{cm^3}$; $J_m$=23 meV, $D$=1.6 meV, $K$=0.17 meV, $\alpha=0.003$ \cite{Tsai2020nature}.}
\label{Fig:Simulation}
\end{figure}

In this Letter, we theoretically show how a kagome noncollinear antiferromagnet can be manipulated by the intrinsic spin-orbit torques from the spin accumulations induced by a pure electric current, following the microscopic mechanism below [Fig. \ref{Fig:Mn3Sn}(b-d)].


\begin{eqnarray}
\begin{matrix}
\mathrm{Pure} &   & \mathrm{Spin} &  & \mathrm{Intrinsic}  & & \mathrm{Magnetic} \\
\mathrm{electric} & \Rightarrow  & \mathrm{accumu}  & \Rightarrow  &  \mathrm{spin}-\mathrm{orbit} & \Rightarrow & \mathrm{dynamics} \\
\mathrm{current}  &    & \mathrm{-lation}  &  & \mathrm{torque} & & \mathrm{of\ moments}
\end{matrix}\nonumber
\end{eqnarray}
We self-consistently calculate the electronic structure, magnetic
structure, spin accumulations, and intrinsic spin-orbit torques (Fig. \ref{Fig:flowchart}). We point out
the significance of parity symmetry breaking in deterministic switching (Fig. \ref{Fig:parity}). Our calculation can yield a critical current density comparable with those in the experiments (Fig. \ref{Fig:Simulation}), when considering the boost from the out-of-plane magnetic dynamics induced by the current-driven spin accumulation on individual magnetic moments.
Our description of the magnetic dynamics of the kagome noncollinear antiferromagnet will be helpful for future applications.

\textcolor{blue}{ \emph{Intrinsic spin-orbit torques.}} -  First, we illustrate the spin-orbit torques from the spin accumulations induced by the pure electric current. The noncollinear antiferromagnet $\mathrm{Mn_3Sn}$ is formed by stacked bilayers along the $c$ ($y$ here) axis (Fig. \ref{Fig:Mn3Sn}). One layer in the bilayer is the inversion of the other, composed of a Mn kagome lattice and a Sn hexagonal lattice. In the unit cell of the Mn lattice, the magnetic moments of the three Mn atoms are rotated with respect to each other by nearly 120$^\circ$, forming a noncollinear antiferromagnetic structure (Sec. S1 of \cite{Supp}).
The unit moment $\mathbf{m}_{a}$ on Mn atom $a$ is governed by \cite{Liu2017prl}
\begin{eqnarray}\label{MH}
H_m&=&2\sum_{ab}\left[J_m\mathbf{m}_{a}\cdot\mathbf{m}_{b}+D\mathbf{\hat{y}}\cdot(\mathbf{m}_{a}\times\mathbf{m}_{b})\right]\nonumber\\
    &&-K\sum_{a}(\hat{\mathbf{e}}_a\cdot \mathbf{m}_{a})^2,
\end{eqnarray}
where $a,b\in\{1,2,3\}$ index the three Mn atoms in the unit cell, the first summation runs over the combinations $ab\in\{12,23,31\}$. The nearest neighbor Heisenberg interaction $J_m$ and Dzyaloshinskii-Moriya interaction $D$ describe a perfect triangular noncollinear antiferromagnetic structure with exactly 120$^\circ$ between the Mn magnetic moments. The anisotropy $K$ term
breaks the in-plane $U(1)$ symmetry and leads to six stable positions to which the magnetic structure tends to relax \cite{Liu2017prl} (see Fig. \ref{Fig:Simulation}), $\hat{\mathbf{e}}_a$ represents the anisotropic axis for $\mathbf{m}_a$. The injected current has to exert enough spin-orbit torques to overcome the six stable positions for switching.

The dynamics of the magnetic moments $\mathbf{m}_{a}$ is described by the Landau-Lifshitz-Gilbert equation (Sec. S2A of Ref. \cite{Supp})
\begin{eqnarray}\label{Eq:LLG}  (1+\alpha^2)\dot{\mathbf{m}}_{a}&=&\frac{|\gamma|}{M_S} \mathbf{m}_{a}\times \frac{\delta H_m}{\delta\mathbf{m}_{a}}+\mathbf{T}_{a}\nonumber\\
&&+\alpha \mathbf{m}_{a}\times\left(\frac{|\gamma|}{M_S} \mathbf{m}_{a}\times \frac{\delta H_m}{\delta\mathbf{m}_{a}}+\mathbf{T}_{a}\right),
\end{eqnarray}
where $\gamma (<0)$ is the gyromagnetic ratio of the electron, $\alpha$ denotes the Gilbert damping coefficient,
the saturation moment of a Mn atom $M_S=3\mu_B$ with $\mu_B$ the Bohr magneton,
and $\mathbf{T}_{a}$ is the spin-orbit torques induced by the electric current. Different from the extrinsic damp-like spin-orbit torques \cite{Liu2011prl,Slonczewski1996JMMM}
$\mathbf{T}_{a}\sim \mathbf{m}_{a}\times(\mathbf{m}_{a}\times \mathbf{s})$ caused by externally injected spin currents with a uniform spin $\mathbf{s}$,
we can show that the exchange interactions between the Mn moments and current-induced spin accumulations (Sec. S2B of Ref. \cite{Supp}) naturally give the intrinsic field-like spin-orbit torques \cite{Haney2008JMMM,Manchon2019handbook}
\begin{eqnarray}
  \mathbf{T}_{a}= \frac{2|\gamma|J_{sd}V_U}{M_S\hbar}
  \mathbf{m}_{a}\times \tilde{\mathbf{s}}_a, \label{SOT-H}
\end{eqnarray}
where the torques from the divergence of spin current and spin-orbit coupling \cite{Manchon2019handbook} have been taken into account, $J_{sd}$ is the exchange interaction between the Mn moments and itinerant electron spins, $V_U$ is the volume of the unit cell, and inter-layer antiferromagnetic interactions (e.g., 1-2$'$ and 1-3$'$) tend to synchronize the diagonal moments (i.e., 1 and 1$'$) on two layers, so we use the averaged local spin accumulation density $\tilde{\mathbf{s}}_a=(\mathbf{s}_a+\mathbf{s}_{a'})/2$. With the help of the linear-response theory \cite{Mahan00book} (derivations in Sec. S3 of Ref. \cite{Supp}), the local spin accumulation density on Mn atom $a$
induced by a pure electric current are found to have two parts ($d$ for diagonal and $od$ for off-diagonal matrix elements of the operators $v$ and $\sigma$)
\begin{eqnarray}
\mathbf{s}_a^d&=&\frac{e\hbar}{2V}\tau \sum_{\nu,\mathbf{k}}\frac{\partial f_\nu}{\partial\epsilon_\nu}(\mathbf{E}\cdot \mathbf{v}_{\nu\nu})\bm{\sigma}_{\nu \nu}^{a}, \label{Eq:Kubo-d} \\
  \mathbf{s}_a^{od}&=&\frac{e\hbar^2}{2V} \sum_{\mu\neq \nu,\mathbf{k}} (f_\mu-f_\nu)\mathrm{Im}\left[\frac{(\mathbf{E}\cdot \mathbf{v}_{\mu \nu})\bm{\sigma}_{\nu\mu}^{a}}{(\epsilon_\mu-\epsilon_\nu)^2}\right] \label{Eq:Kubo-od},
\end{eqnarray}
where the current density $\mathbf{j}$ injected along an arbitrary direction enters as an electric field
\begin{eqnarray}\label{Eq:current}
\mathbf{E}=\frac{m^*}{ e^2 n_{3D} \tau } \mathbf{j},
\end{eqnarray}
$e=-1.6\times 10^{-19}$ Coulomb, $m^*$ is the effective mass, $n_{3D}$ is the carrier density, $\tau$ is the relaxation time. Under the driving electric current, the system will enter a nonequilibrium static state, in a time scale described by $\tau$ (usually 1-100 picoseconds) in the current-spin correlation Eq. (4). $\tau$ is much shorter than the nanosecond timescale of switching in Fig. \ref{Fig:Simulation}, where the spin density is relaxed by the torque $\mathbf{T}_a$ in Eq. (\ref{SOT-H}) to transfer angular momenta to the magnetic structure. $V$ is the volume of the system, $f$ is the Fermi-Dirac distribution function, $\bm{\sigma}^{a}=(\sigma_x^a,\sigma_y^a,\sigma_z^a)$ are the Pauli matrices for the local spin density on Mn atom $a$, $\epsilon$ and $v$ are the energy and group velocity operators, $\nu$ and $\mu$ label the energy bands, described by an $s$-$d$ model (Sec. S4 of Ref. \cite{Supp})
\begin{eqnarray}\label{Eq:sd}
H_{sd}&=&J_{sd}\sum_{i,a,n,n'}\mathbf{m}_{a}\cdot\bm{\sigma}_{nn'}c^{\dagger}_{ian}c^{}_{ian'}+H_{e},
\end{eqnarray}
where $J_{sd}$ is the exchange interaction between the Mn moments ($a\in\{1,2,3,1',2',3'\}$) and spins of itinerant electrons ($n\in\{\uparrow,\downarrow\}$),
and $i$ indexes the unit cell.
{ The itinerant electron part $H_e$ needs to include parity symmetry breaking, as illustrated below.}

\textcolor{blue}{ \emph{Requirement of parity symmetry breaking.}} - The parity operation transforms the current (a radial vector) to the opposite direction while leaving the Mn moments (axial vectors) invariant, which means that an opposite current could also lead to the same final state of the Mn moments, so there is no one-to-one deterministic relation between the current direction and magnetizations of the Mn moments under parity symmetry (Fig. \ref{Fig:parity}). This can also be seen from a symmetry analysis of Eqs. (\ref{Eq:Kubo-d}) and (\ref{Eq:Kubo-od}) (Sec. S5 of Ref. \cite{Supp}).
Required by parity symmetry breaking, we figure out all possible linear-momentum Hamiltonian terms that break parity symmetry, as shown in Fig. \ref{Fig:parity}. To construct the itinerant electron Hamiltonian $H_e$, we choose to add $p_y\sigma_y$ term with magnitude $\Delta_p^\mathrm{I}$ or $p_y$ term with magnitude $\Delta_p^\mathrm{II}$ to break parity symmetry in the previously-proposed three-dimensional (3D) Weyl model \cite{Batista2020prb}. Its tight-binding model reads
\begin{eqnarray}\label{Eq:Weyl}
H_{e}^{W}&=&\sum_{\langle ia,jb\rangle \parallel}c^{\dagger}_{ia}t_{ia,jb}c^{}_{jb}+\sum_{\langle ia,jb\rangle \perp}c^{\dagger}_{ia}t'^{}_{ia,jb}c^{}_{jb}\nonumber\\
  &\ &+\sum_{\langle ia,jb\rangle}c^{\dagger}_{ia}\mathbf{i}(\Delta_p^\mathrm{I}\sigma_y+\Delta_p^\mathrm{II})\mathrm{sgn}(\hat{\mathbf{y}}\cdot\hat{\mathbf{d}}_{ia,jb})c^{}_{jb},
\end{eqnarray}
where $c_{ia}=(c_{ia\uparrow},c_{ia\downarrow})^{T}$, $\parallel$ and $\perp$ distinguish the intra-layer and inter-layer nearest neighbor sites, $t_{ia,jb}$ = $t \exp[\pm \mathbf{i} \alpha_1\sigma_y/2]$, with $+$ for $ab\in\{$12, 23, 31, $1'2'$, $2'3'$, $3'1'$$\}$ and $-$ for $ab\in\{$21, 31, 13, $2'1'$, $3'2'$, $1'3'$$\}$, $t'_{ia,jb}=t \exp[-\frac{\mathbf{i}}{2}\mathrm{sgn}(\hat{\mathbf{d}}_{ia,jb}\cdot \hat{\mathbf{y}}) \alpha_2 (\cos \theta\hat{\mathbf{d}}^{\parallel}_{ia,jb}+\sin \theta\hat{\mathbf{y}})\cdot\bm{\sigma}]$. $\alpha_1$, $\alpha_2$, and $\theta$ are the model parameters. With the help of symmetry analysis, we find that for $\Delta_p^{\mathrm{I}}\neq 0$, the non-zero diagonal term of spin accumulations are in-plane and non-zero off-diagonal term of spin accumulations are out of plane; for $\Delta_p^{\mathrm{II}}\neq 0$, the non-zero diagonal term of spin accumulations are out of plane and non-zero off-diagonal term of spin accumulations are in plane. Another choice to break parity symmetry is a 2D Rashba model with $p_x\sigma_z-p_z\sigma_x$ term, as compared in Sec. S5 of Ref. \cite{Supp}.

\begin{table}[t]
\caption{Diagonal [Eq. (\ref{Eq:Kubo-d})] and off-diagonal [Eq. (\ref{Eq:Kubo-od})] spin accumulations on the diagonal lattice sites $(a,a') \in\{(1,1'), (2,2'), (3,3')\}$, { constrained by the symmetries of the model in Eq. (\ref{Eq:Weyl}) with two different parity symmetry breaking terms $\Delta_p^{\mathrm{I}}$ and $\Delta_p^{\mathrm{II}}$.} Here, $\parallel\in \{x,z\}$ and $\perp=y$ mean in and out of the kagome planes, respectively.  $P$ and $T$ are the inversion and time-reversal operations, respectively, $C_{2y}$ is the two-fold rotation around the $y$ axis, and $\text{\boldmath$\tau$}_y$ is the translation along the $y$ axis for half lattice constant $c/2$.}
\label{tab1}
\begin{ruledtabular}
\begin{tabular}
{cccc}
&Symmetry&Diagonal & Off-diagonal
\\
\hline
$\Delta_p^{\mathrm{I}}\neq 0$,   &$T\{C_{2y}|\text{\boldmath$\tau$}_y\}$&$(\mathbf{s}_{d}^{a}+\mathbf{s}_{d}^{a'})_{\parallel}\neq 0$ & $(\mathbf{s}_{od}^{a}+\mathbf{s}_{od}^{a'})_{\parallel}=0$
\\
$\Delta_p^{\mathrm{II}}=0$ &&$(\mathbf{s}_{d}^{a}+\mathbf{s}_{d}^{a'})_{\perp}=0$ & $(\mathbf{s}_{od}^{a}+\mathbf{s}_{od}^{a'})_{\perp}\neq 0$
\\
\hline
$\Delta_p^{\mathrm{I}}= 0$,  &$T\{PC_{2y}|\text{\boldmath$\tau$}_y\}$&$(\mathbf{s}_{d}^{a}+\mathbf{s}_{d}^{a'})_{\parallel}=0$
& $(\mathbf{s}_{od}^{a}+\mathbf{s}_{od}^{a'})_{\parallel}\neq 0$
\\
$\Delta_p^{\mathrm{II}}\neq 0$
&&$(\mathbf{s}_{d}^{a}+\mathbf{s}_{d}^{a'})_{\perp}\neq 0$ & $(\mathbf{s}_{od}^{a}+\mathbf{s}_{od}^{a'})_{\perp}=0$
\\
\end{tabular}
\end{ruledtabular}
\end{table}

\textcolor{blue}{ \emph{Complete simulation of magnetic dynamics.}} - Following the steps in Fig. \ref{Fig:flowchart}),  we perform numerical simulations to show how the pure electric current switches the magnetic structure of the kagome non-collinear antiferromagnet. The results for one of the Weyl model ($\Delta_p^\mathrm{I}\neq 0$, $\Delta_p^\mathrm{II}= 0$) are shown in Fig. \ref{Fig:Simulation} (a), measured by the magnetic moment angle of one of the Mn atoms $\varphi_2$. As the current along the $+z$ ($-z$) direction is turned on and above a critical current density, the magnetic structure can be fully polarized to a metastable position at $\varphi_2=\pi/2$ ($\varphi_2=-\pi/2$). As the current is turned off, the magnetic structure relaxes to the nearest one of the six stable positions. More simulations for the other Weyl model and Rashba model can be found in Fig. S8 of Sec. S4C in Ref. \cite{Supp}.

Our simulations yield a low critical current density comparable with those in the experiments \cite{Deng2022nsr}, for two reasons. 
First, we reveal a boost from the out-of-plane magnetic dynamics, similar to a case in the collinear antiferromagnets \cite{Gomonay18np}. According to our simulations, the torque is dominantly contributed by the scenario in Fig. \ref{Fig:Mn3Sn}(b), where, taking Mn 2 for example ($\mathbf{m}$), its current-induced in-plane spin accumulation $\mathbf{s}_{||}$ can induce an out-of-plane component $m_{\perp}$ under the field-like torque $\mathbf{m} \times \mathbf{s}_{||}$, as shown by Fig. \ref{Fig:Mn3Sn}(b). For Mn 2, its antiferromagnetic interactions with other two Mn moments [i.e., $(|\gamma|/M_S) \mathbf{m}_a\times (\delta H_m/\delta \mathbf{m}_{a})$ in Eq. (\ref{Eq:LLG})] act like an effective field $\mathbf{H}_\mathrm{AFM}$ and $m_{\perp} \times \mathbf{H}_\mathrm{AFM}$ gives the torque in the kagome plane, as shown by the orange $\otimes$ in Fig. \ref{Fig:Mn3Sn}(d) and orange arrow in Fig. \ref{Fig:Mn3Sn}(b). This out-of-plane dynamics is efficient, because  
$\mathbf{H}_\mathrm{AFM}\sim 2J_m/M_s\sim $ 264 Tesla, is pretty strong. This out-of-plane magnetic dynamics also occurs to Mn 1 and Mn 3. When the total out-of-plane moments of three Mn atoms $(\mathbf{m}_1+\mathbf{m}_2+\mathbf{m}_3)_\perp > 0$ ($<0$), the entire magnetic structure rotates counter-clockwise (clockwise) among stable and metastable positions. Fig. \ref{Fig:Simulation}(b) shows that a peak of $(\mathbf{m}_1+\mathbf{m}_2+\mathbf{m}_3)_\perp $
always emerges along with a transition in Fig. \ref{Fig:Simulation}(a), indicating the significance of the boost from the out-of-plane magnetic dynamics. 
By contrast, although the damp-like torque $\mathbf{m}\times (\mathbf{m}\times \mathbf{s}_{\perp})$ from out-of-plane spin accumulation $\mathbf{s}_{\perp}$ in Fig. \ref{Fig:Mn3Sn}(c) can also induce the out-of-plane moment, it is much smaller because of the small damping constant $\alpha$ in front (see Sec. S2C in Ref. \cite{Supp}). Second, we assume that the Mn atoms feel different fields locally produced by the spin accumulations in the simulation. In contrast, if we assume that the Mn atoms feel the same field collectively (like the torques from the extrinsically injected spin currents \cite{Tsai2020nature}), the critical current density has to be about three orders larger (see Fig. S13 in Sec. S7B of Ref. \cite{Supp}). 
The sharp difference can be understood as follows. The above-mentioned $(\mathbf{m}_1+\mathbf{m}_2+\mathbf{m}_3)_\perp$ is proportional to  $\mathbf{m}_1\times \mathbf{s}_1+\mathbf{m}_2\times \mathbf{s}_2+\mathbf{m}_3\times \mathbf{s}_3$ under the different-field assumption, and
$(\mathbf{m}_1+\mathbf{m}_2+\mathbf{m}_3)\times \mathbf{s}$
in the same-field assumption, where $ \mathbf{s} = \mathbf{s}_1+ \mathbf{s}_2+\mathbf{s}_3$, comparable with each of $\mathbf{s}_a$ (see Fig. S9 of Ref. \cite{Supp}).
For the noncollinear 120$^\circ$ texture of Mn$_3$Sn,
$\mathbf{m}_1,\mathbf{m}_2,\mathbf{m}_3$ nearly cancel with each other, and $\mathbf{m}_1+\mathbf{m}_2+\mathbf{m}_3$ is about 0.01$\mu_B$ \cite{Tsai2020nature}, three orders smaller than that of $\mathbf{m}_i\sim 3\mu_B $, so the critical current density under the same-field assumption is about three orders larger.

Our simulations also reveal many microscopic details. For the current density of $54\times 10^6$ in Fig. 4(a), after turning off the current, it takes a few ns to transit to another state. This feature appears when the magnetic structure is very close to a metastable position (e.g., $\varphi_2=\pi/2$) at the middle of two neighbor stable positions ($\varphi_2=\pi/3$ and $2\pi/3$). The argument is, at exactly the metastable position, there is no preference of falling to any one the two equivalent neighbor stable positions, so it takes a longer time to break the detailed balance to escape from the vicinity of the metastable position. To illustrate this, we present the simulations starting from slightly below, extremely close to, and slightly above a metastable position (Fig. S15 in Ref. \cite{Supp}). 

\begin{acknowledgements}
$^\dag$Y.C. and Z.Z.Du contributed to the work equally.
We thank helpful discussions with Kaiyou Wang. This work was supported by the National Key R\&D Program of China (2022YFA1403700), Innovation Program for Quantum Science and Technology (2021ZD0302400), the National Natural Science Foundation of China (11925402 and 12004157), Guangdong province (2020KCXTD001 and 2016ZT06D348),
and the Science, Technology and Innovation Commission of Shenzhen Municipality (ZDSYS20170303165926217, JAY20170412152620376, and KYTDPT20181011104202253). The numerical calculations were supported by Center for Computational Science and Engineering of SUSTech.
\end{acknowledgements}

\bibliographystyle{apsrev4-1-etal-title_6authors}
\bibliography{refs-transport,refs-Mn3Sn}

\end{document}